\documentclass[11p]{article}

\usepackage{tablefootnote}

\usepackage{natbib}

\date{}

\begin{document}

\title{Research Methods in Computer Science: The Challenges and Issues}

\author{Hossein Hassani \thanks{hossein.hassani@stu.ssst.edu.ba} \thanks{hosseinh@ukh.edu.krd}
\thanks{Hossein Hassani is a lecturer at the University of Kurdistan Hewl\^er and a visiting lecturer at the Sarajevo School of Science and Engineering}}

\maketitle

\begin{abstract}
	 Research methods are essential parts in conducting any research project. Although they have been theorized and summarized based on best practices, every field of science requires an adaptation of the overall approaches to perform research activities. In addition, any specific research needs a particular adjustment to the generalized approach and specializing them to suit the project in hand. However, unlike most well-established science disciplines, computing research is not supported by well-defined, globally accepted methods. This is because of its infancy and ambiguity in its definition, on one hand, and its extensive coverage and overlap with other fields, on the other hand. This article discusses the research methods in science and engineering in general and in computing in particular. It shows that despite several special parameters that make research in computing rather unique, it still follows the same steps that any other scientific research would do. The article also shows the particularities that researchers need to consider when they conduct research in this field.

\end{abstract}

\section{Introduction}
\label{intro}
	
Dictionaries define research as ``the systematic investigation into and study of materials and sources in order to establish facts and reach new conclusions. \citep{fowler2011concise}'' Scholars, for example, \cite{depoy2015introduction} suggest that the definition should be elaborated and provided in the context of the research area and the related branch of science. Therefore, after arguing on their opinion that the broad definition of the research would not be of practical benefit in research, they provide their definition for their field of interest, which in this case is humanities, as ``multiple, systematic strategies to generate knowledge about human behavior, human experience, and human environments in which thinking and action processes of researcher are clearly specified so that they are logical, understandable, confirmable, and useful'' \citep{depoy2015introduction}. They also refer to research as ``a multiple, systematic strategy to generate knowledge about human behaviour, human experience, and human environments,'' which the researcher conducts it through applying and following an explicit process \citep{depoy2015introduction}. Again, mentioning that the term "research" has been defined in different ways, \cite{punch2000developing} redefines it as ``an oraganized, systematic and logical process of inquiry, using empirical information to answer questions (or test hypothesis). \citep{punch2000developing}'' By hypothesis, we mean an statement that can be either true or false, which researchers try to find out through conducting their research \citep{depoy2015introduction}. The mentioned definitions, and others as they can be found by consulting research methods literature, almost with the same ingredients, suggest that research have, at least, three concepts in common:  (a) to devise a research question, (b)to suggest and follow an approach or method to address/investigate/solve this question, and (c) to provide the results after the mentioned methods applied.

\par Although the above definitions are applicable to the research in all categories of science and humanities, the characteristics of each field requires particular adaptation of the concepts through understanding the nature of the specific research. This article discusses the research methods and methodology in science, in general, and focuses on the concept in the context of computing and computer science, in particular.

\par In the following sections, first, the different paradigms of research method, in different branches of science (and humanity) are discussed; next, the approach of natural sciences to research is presented; afterwards, the main paradigm(s) that lay(s) the foundation of research approaches in computing and computer science are discussed; next, the theoretical foundation of research methodology in the context of computing and computer science is presented; finally, the last section summarizes the discussion and provides the conclusion. \\

\section{Research Paradigms}
\label{paradigm}

\par Generalizing the results of or findings from some experiments in an area in a way that could be applied or used beyond the specific area under investigation is the main goal of research \citep{hammersley2012ethics, de2002analyzing}. From this perspective, the main goal of research remains the same, no matter of in what branch of a science it is conducted. This is also regardless of science categorization. That is, for example, natural science, social science, applied science, behavioral science, and humanities share the same main goal in there research. However, the way that scientists are looking at a phenomenon and raise questions about it significantly differs depending on the branch and category of the science that they are active in. The approaches that they take to solve the related problems and to answer the questions also differ considerably. As a result, several paradigms about research have formed. \cite{grbich2013qualitative} defines paradigms as ``worldviews of beliefs, values, and methods for collecting and interpreting data. \citep[p. 5]{grbich2013qualitative}'' \cite{denicolo2012developing} define it as ``a basic set of beliefs, views, values and assumptions that guide action and include the researcher's epistemological, ontlogoical and methodological premises \citep{denicolo2012developing}'' , in which ``epistemology'' refers to the theory of the formation of knowledge \footnote{A more broader definition of epistemology has been given in \citep[pp. 237-290]{wray2002knowledge}. Although the concept is mainly discussed from social sciences perspective, it provides a thorough insight into the concept from several viewpoints and by different writers through nine chapters.} and ``ontolgoy'' is the study of the things and the nature of being \footnote{Ontology is a concept of philosophy. The word Ontology in the context of computer and informations sciences has been specialized. This specialization has been provided in \citep{stanford2015ontology}, which refers the definition of the word by \cite{gruber1993translation}. This specialization has been updated in 2009 \citep{gruber2009ontology}. In addition, a more detailed discussion on the subject in the context of computing can be found in \citep{uschold1996ontology}.}.

\par Although similar enough to draw a general perspective, the categorization of research paradigms varies among the scholars (see \citep{denicolo2012developing, grbich2013qualitative, punch2000developing}). For example, \cite{denicolo2012developing} categorize the paradigms of research as \textbf{positivism}, \textbf{post-positivism}, \textbf{constructivism}, and \textbf{critical theory} \citep{denicolo2012developing}, while \cite{grbich2013qualitative} presents it as \textbf{realism/postpositivism}, \textbf{critical theory}, \textbf{interpretivism/constructionism}, \textbf{postmodernism and poststructuralism}, and \textbf{mixed/multiple methods}.

\par Table  1 shows a brief description of different research paradigms. The definitions are according to \cite{denicolo2012developing}; other resource such as \cite{mackenzie2006research} also provide these categorization with some more details.

\begin{table}[!h]
\label{tbl:Prdgm}
\begin{center}
\begin{tabular}{|p{5.25cm}|p{5.25cm}|}
\hline
Positivism / Post-positivism & Constructivism / Critical Theory \\ \hline
\begin{itemize}
\item Nomothetic \tablefootnote{Nomothetic means generalizing ideas by finding common base and to extract abstracts from different related or correlated phenomenon.}
\begin{itemize}
\item abstraction
\item law generation
\item generalization (universalization)
\item investigating and explaining relationships (causal or correlation) between phenomena 
\item finding and manipulating related variables
\item evaluation of hypothesis
\end{itemize}
\end{itemize}
\begin{itemize}
\item Deductive
\end{itemize}
\begin{itemize}
\item Reductionist
\end{itemize}
\begin{itemize}
\item Objectivist
\end{itemize}
\begin{itemize}
\item Data collection
\begin{itemize}
\item mainly quantitative
\end{itemize}
\end{itemize}
\begin{itemize}
\item Dominance
\begin{itemize}
\item natural science
\item life science
\end{itemize}
\end{itemize}  & 
\begin{itemize}
\item Idiographic (mainly) \tablefootnote{Idiographic approach means to study things from individuals view or the groups of people perspective. This is of very importance in social sciences. However, unless a computer scientist is not involved in an interdisciplinary research that has some social or life sciences concern, more details do not seem to be relevant.}
\begin{itemize}
\item specializing
\item unique understanding
\item individualization
\item defining research questions 
\item finding answers to the research questions
\vspace*{5\baselineskip}
\end{itemize}
\end{itemize}
\begin{itemize}
\item Phenomenological
\end{itemize}
\begin{itemize}
\item Interpretivist
\end{itemize}
\begin{itemize}
\item Subjectivist
\end{itemize}
\begin{itemize}
\item Data collection
\begin{itemize}
\item mainly qualitative
\end{itemize}
\end{itemize}
\begin{itemize}
\item Dominance
\begin{itemize}
\item social science
\item humanities
\end{itemize}
\end{itemize} \\ \hline
\end{tabular}
\end{center}
\caption{Current research paradigms} 
\end{table} 

\par Despite the differences that can be seen in different research paradigms, in many situation a combination of approaches that theses paradigms suggest would serve the research design much better than a single one. Therefore, a mix/multi-methods paradigm which is obtained from an amalgamation of different paradigms and their related methods has received more attention in recent studies \citep{johnson2007toward}.  In fact, \cite{ramesh2004research} show that research in computing have been conducted according to broad range of paradigms and approaches.

\par However, adapting a field specific approach is a necessary step in every branch of science. To that extent, computer science as a fairly new discipline, suffers from ``lack of identity'', although it combines the experience of its main roots, namely, mathematics and engineering, \citep{demeyer2011research}. In contrast, there are others who believe that ``computer science is a well-established discipline'' that it has all it needs to be considered as any other sciences with a long history \citep{ramesh2004research}. In either case, the mentioned main roots affect the formation of computing research paradigm. That is, a positivism/realism which is the main paradigm of natural and life sciences, is applied to computing as well \citep{denicolo2012developing}.

\section{Research Methods versus Research Methodology}
\label{methodology}

Despite uncountable books, articles, and discussions about research \textbf{method} and \textbf{methodology}, finding a straightforward definition seems not to be easy. In fact, the issue is with interpretation of methodology rather than method. Some scholars, such as \cite{clough2012student}, for example, have tried to discuss the terms in more detail. However, they do not seem to be giving a clear-cut and concise differentiation between the two terms. Similarly, \cite{mcgregor2010paradigm} have provided some more explanation, mentioning that ``the word methodology comprises two nouns: method and ology, which means a branch of knowledge''. This is also similar to what can be found in \citep{Dictionary.com2015}. Differently, \citep{merriam2015dic} sets origin of methodology to ``New Latin methodologia, from Latin methodus + -logia -logy'' and dates it back to 1800. Yet another article by \citep{lehaney1994methodology} focuses on the usage of the term in the specific context and tries to enlighten the readers about the confusions around this particular usage. But despite all efforts, it seems that ``methodology'' continues to remain as a confused term in the research community. Particularly, it has been taken for granted in many ways and has interchangeably been used alongside ``method'' in many resources. However, in this article we try to clearly differentiate these two in a way that follows.

\par We consider ``method'', in the context of research, as an approach, procedure, and guidelines that are used in conducting a research. A method might require different tools, instruments, equipments, and such. Whereas we consider ``methodology'' as a scientific approach that investigates, compares, contrasts, and explains the different ways that a research could be conducted alongside different methods that could be used in these processes. That is, methodology discusses the alternative approaches and methods to tackle the research problem. It discusses the advantages/disadvantages, properness/improperness, feasibility, practicality, ethical issues, and such parameters for the approaches to do the research. Throughout its discussion, the methodology, as a main ingredient of the research, clarifies why a particular approach has been taken to address the ``research question(s)'' and how this approach would be implemented.
     
\par Based on what is mentioned, the research methodology should reflect on the nature of the research and help the researcher to tackle the research area, properly. For this purpose, the researcher should find out, through theoretical/factual discussions, the research category and the paradigm, which better show the characteristics of the research and serve the research to be conducted more properly. For this, \cite{baban2009research} categorizes research based on three main themes, which have been summarized as below: 

\begin{enumerate}
\item The application of the research study

\begin{itemize}
\item \textbf{Pure research} – It aims in discovering new knowledge without expecting an instant affect on the current situation of the field.
\item \textbf{Applied research} – It aims in solving a specific problem, which is currently the concern of the field.
\end{itemize}

\item The objectives in undertaking the research

\begin{itemize}
\item \textbf{Descriptive} research – It aims in explaining the situation and characteristic of a specific problem in order to benefit from it in other research.
\item \textbf{Exploratory} research – It aims in finding proper information in the area within which researcher cannot find previous information in order to build a profound hypothesis.
\item \textbf{Correlational} research – It aims in discovering the correlations among different variables of problem area in order to recognize the impacts of a phenomenon.
\item \textbf{Explanatory} research – It aims in explaining the reasons behind the characteristics of a phenomenon (answering to the why) or how the characteristics of a phenomenon forming it. 
\item \textbf{Analytical} research - It can be considered as an extension to the descriptive research because it does not stay at the description level, and moves beyond that to discover the reasons behind a problem or the behavior of a phenomenon.
\end{itemize}

\item The type of information sought

\begin{itemize}
\item \textbf{Positivism} - see Table 1.
\item \textbf{Phenomenological} - see Table 1. 
\end{itemize}

\end{enumerate}

\par To set a proper paradigm and to suggest well-suited methods that could best serve the research purpose are paramount to the research. \cite[pp.~28-29~]{baban2009research}
discusses both quantitative and qualitative approach based on certain assumptions that researchers may make. These assumptions are ontological assumptions, epistemological assumptions, axiological assumptions, rhetorical assumptions, and methodological assumptions. Others such as \cite{saunders2007research} also give similar perspectives in this regard. Although it seems that these are different opinions, but the conclusions are similar, and the main differences remain in the way that the ideas are presented.

\section{Research in Computing} 
\label{computing}

\cite{dodig2002scientific}, quoting Dijkstra, mentions that computer science departments, have, under external pressures, underemphasized the ``science'' aspects of the knowledge area in favor of ``computer'' that is a tool not a science; this implies that, for example, the surgeons call surgery a ``knife science'', or having ``car engineering'', ``train engineering'', or rather ``car engine engineering''! This approach focuses on computers as a tool rather than appreciation of theoretical aspects and abstract elements of this science, such as mathematics and logic, one the one hand, and undermines its engineering elements as a necessary part, particularly, in software engineering, on the other hand \citep{dodig2002scientific}. Similarly, long while ago in 1970s, \cite{newell1976computer} stated that ``Computer science is an empirical discipline. We would have called it an experimental science, but like astronomy, economics, and geology, some of its unique forms of observation and experience do not fit a narrow stereotype of the experimental method \citep[p. 14]{newell1976computer}.'' \cite{dodig2002scientific} also refers to the three fundamental recurring concepts of computing, which are (a) conceptual and formal models, (b) levels of abstraction, and (c) efficiency.

\par So what is computing? How we can define it? To answer this questions is not as easy as it seems to be at the first glance \citep{snyder1994academic}. We have observed this confusion among applicants who are interested in computing, but they do not know which sector of computing should they choose, because the difference is not made clear for them. However, with regard to undergraduate study in computing there is a consensus among the majority of academics on differences and commonalities among different sectors of computing (for example, see \citep{CS2013, SE2014draft, pyster2009graduate, topi2010curriculum}).

\par According to \cite{CS2013}, ``omputing is a broad field that connects to and draws from many disciplines, including mathematics, electrical engineering, psychology, statistics, fine arts, linguistics, and physical and life sciences.'' Speaking about its past and current situation, \cite{denning2013science} state: ``computing began as science, morphed into engineering for 30 years while it developed technology, and then entered a science renaissance about 20 years ago. Although computing had subfields that demonstrated the ideals of science, computing as a whole has only recently begun to embrace those ideals. Some new subfields such as network science, network social science, design science, and Web science, are still struggling to establish their credibility as sciences.'' \citep[p.~32~]{denning2013science}
	
\par We discussed the research methods and methodology in the context of science and humanities. But how this applies to computing? \cite{milner1986computing}, in a fascinating inaugural lecture to the opening of the Laboratory for Foundations of Computer Science at the University of Edinburgh, has provided answers to a question that seems to be still valid to many people, after some 30 years passed since the lecture was given. That is, ``Is Computing an Experimental Science?''. In fact, there is yet a more rudimentary question: Is Computing a Science? or even: Is Computer Science a Science? It must have been taken for granted that the answers to these questions are rather a simple ``yes!'', however, this is not the case \citep{dodig2002scientific}, and unless we are able to include computing and computer science as valid members of science, we are not able to apply scientific approaches to their research. This, reminds me of a colleague's saying, an academic in humanity, who was showing his full surprise about how could people in computing, particularly in software engineering, receive a PhD and call themselves a ``Doctor of Philosophy'' in a subject that, in his opinion, neither was a science nor engineering!

\par Although the phrase ``Computing is a science.'' seems to be an axiom, for some scholars it is not. National Academy of Science elaborated on this matter in 1992. It refers to both scientific and engineering aspects of computing and relates the first to the mathematical and engineering models, based on theory and abstraction, whereas relates the second to the practical application, based on abstraction and design \citep{nap1992computingfuture}. They also compare the object of study in computer science with other branches of science. For example, if the object of research in physics is atom, or in biology is a cell, then ``focus on information, the ways of presenting information, and on the machines and systems that perform these tasks'' are the objects of study in computer science \citep{nap1992computingfuture}.

\par Regardless of significant advancement in computing, and no matter if computers are becoming a central player in almost every aspect of life in the new millennium, some scientists still believe that ``computing science is an immature discipline.'' \cite{johnson2006research}~argues on the issues of the advancement in computing technology and academic research. From his perspective, although using hermeneutics in requirement analysis, and mathematical models to specify and verify complex systems, for example, has been beneficial to the computing research, ``lack of any agreed research framework reflects the strength and vitality of computing science''. As a result, \citep{johnson2006research}~encourages computing researchers contemplate on the various aspects of the research methods they adopt and critically incorporate them into their own research. He says: \\ ``Too often, MSc and PhD theses slavishly follow empirical or formal proof techniques without questioning the suitability of those approaches. For example, the hermeneutic tradition has delivered results that ignore the constraints of time and money on commercial system development. Formal methods research has produced results that abstract so far away from the problem domain that they cannot be applied or validated. The tragedy is that unless we begin to recognise these failures then we will continue to borrow flawed research methods from other disciplines \citep{johnson2006research}.'' 

\par As a result, research in computing might be of theoretical or experimental nature or a combination of them; it appreciates different paradigmatic views, and utilizes best suited tools and approaches from both quantitative and qualitative methods. One could argue that every other science would do the same, so what is the difference? Well, this article does not intend to make a research in computing a different ``thing'', rather it argues for the similarity of research in computing and other branches of science, while it recognizing its especial characteristics that makes it unique as any other branch of science.

\section{Experimental Computer Science versus Theoretical Computer Science}
\label{experim}

An analogy between research an onion might make the concept more understandable. Some sources have referred to this a ``research onion'' (see \citep[p. 102]{saunders2007research}). Although this multitude layers that are overlapping each other might seem complicated, they are very helpful in the discussion and the design of the research methodology. Particularly, in computer science, to understand whether the research in hand is a theoretical research (sometime it is called basic research \citep{kendal2015howto, saunders2007research}) or experimental one, is a key question which significantly affects the methodology of the research. Now, going back to the questions that \citep{milner1986computing} asked (see \ref{computing}), he clearly showed that he preferred to have a convergence between theory and experiment. He provides and analogy by giving examples of physicists an chemists who improved and refined their theories through taking experiments and suggests the same approach to be taken for computer science, hence he deduces that computer science is as experimental as it is theoretical \citep{milner1986computing}.
     
\par \cite{snyder1994academic} define Experimental Computer Science (ECS) as ``the building of, or the experimentation with or on, nontrivial hardware or software systems.'' In this view, computer science and engineering (CS\&E) should be considered as a whole if one wants to discuss them in the context of experimental research. \cite{johnson2000challenges}, who was awarded the 2010 Knuth Prize, assumes that ``science is the search for the fundamental principles that govern the world around us and explain the phenomena we see'', and then suggests that ``the Theoretical Computer Science (TCS) is the ``science'' underlying the field of computing''. He then concludes that as the computation is basically a discrete logical process, the formal and mathematical nature of TCS is especially appropriate for a science of computing. He also adds that theory is a significant ingredient to not only computing but also for its interdisciplinary characteristics \citep{johnson2000challenges}.

\par Professional bodies also have defined computing and its branches. Below we refer to two quotes. The first one emphasizes the importance of algorithms and their application. This view shows how both theoretical and practical aspects of computing work together and in fact, in some research cases are inseparable. 
``An important part of computing is the ability to select algorithms appropriate to particular purposes and to apply them, recognizing the possibility that no suitable algorithm may exist. This facility relies on understanding the range of algorithms that address an important set of well-defined problems, recognizing their strengths and weaknesses, and their suitability in particular contexts. Efficiency is a pervasive theme throughout this area. \citep[p.~55~]{CS2013}''

\par The second quote shows how computing and computer science need to focus on abstraction, which is in turn a reason for looking in some computer science research as Nomothetic from paradigmatic point of view. 
``Abstraction is a fundamental concept in computer science. A principal approach to computing is to abstract the real world, create a model that can be simulated on a machine. The roots of computer science can be traced to this approach, modeling things such as trajectories of artillery shells and the modeling cryptographic protocols, both of which pushed the development of early computing systems in the early and mid-1940s. \citep[p.~70~]{CS2013}''

\par Finally, there is on aspect of research that is growing steadily among different fields and branches of science and humanities, which is the interdisciplinary characteristics of recent studies. \cite{nap2004interdiscip} define interdisciplinary research as: ``[p. 26]{nap2004interdiscip}Interdisciplinary research (IDR) is a mode of research by teams or individuals that integrates information, data, techniques, tools, perspectives, concepts, and/or theories from two or more disciplines or bodies of specialized knowledge to advance fundamental understanding or to solve problems whose solutions are beyond the scope of a single discipline or area of research practice.'' They also suggest how to evaluate a proposal for its disciplinary coverage. This is how they stated: ``[p.~169]{nap2004interdiscip}Evaluate a proposal to its cell-biology research program by using researchers in cell biology and including a substantial number in chemistry,
physics, computer science, the social sciences, and the humanities as appropriate; this practice would help to ensure disciplinary breadth and reduce bias.'' In fact, many contemporary problems cannot be solved through one aspect of knowledge \citep{nap2004interdiscip}. Computing plays a great role in this aspect of research. As the result, we can expect more and more interdisciplinary, cross-disciplinary, and multi-disciplinary research that is one way or another has utilized computing, or rather has been intertwined with computing. 

\subsection{Research Topics: Examples }

\cite{wilson1952introduction} believes that ``many scientists owe their greatness not to their skill in solving problems but to their wisdom in chossing them.'' He also states that ``the most rewarding work is usually to explore a hithherto untouched field'', which ``are not easy to find today. \cite{wilson1952introduction} also states that '' Although from the day that these have been said, the research field has dramatically changed, but his own ``wisdom'' is more appreciated when one reads his book in the context of computing, a science which was in its infancy at the time. The following quotation from his book is still invaluable, particularly, when one looks at different research questions in computing and one wants to choose a path for the research in this area. ``A research worker in pure science who does not have at all times more problems he would like to solve than he has time and means to investigate them probably is in the wrong business. He may be an excellent experimenter and may have all the qualities required for success in applied science, but he lacks qualities of mind important for pure science. This is not at all to imply that applied science is easier, less demanding, on in any way inferior to pure science; it requires its own special abilities, but they are somewhat different \cite[p.~2~]{wilson1952introduction}.''

\par In the following sections, some sample research topics for both TCS and ECS are listed. The topics have been selected from a collection obtained through using different search engines. For each topic a brief description is provided that shows their main focus area.

\subsubsection{Theoretical Computer Science}

Theoretical Computer Science (TCS), as it was mentioned earlier in this section, encompasses the formal rules, mainly based on mathematics and logic, which are underlying computing science as whole. Therefore, most of the research cases in this field form hypotheses that lead to generalization of findings in order to form a theorem or a formal model, or suggest improvements to the previous formal models and algorithms, or other kinds of theorization that expands the scientific background of this field of computing.

\par For example, below are a list of articles which are related to TCS research:

\begin{itemize}
\item A fast string searching algorithm - This research is about improving a searching algorithm, providing a theoretical analysis of the suggested improvements \citep{boyer1977fast}. 
\item The smallest automation recognizing the subwords of a text - Automata, finite automaton, and deterministic finite automaton (DFA) have been essential parts of theoretical computer science for a long while. This research provides an algorithm to build a smallest partial DFA for a certain problem \citep{blumer1985smallest}.
\item A faster algorithm for testing polynomial representability of functions over finite integer rings - This research is also an improvement to an already devised algorithm in polynomial representability. Reading the article that describes this research, one, at the first glance, would say that is a research in mathematics \citep{guha2015faster}. However, when it is read carefully, the algorithms that have been provided explain why this research has happened in the computer science area.
\item Categorial dependency grammars - Formal grammars are another essential part of TCS that are of different usage in programming languages and other formal language processing in computing. This paper provides an ``abstract theoretical version of sub-commutative'' categrial dependency grammars \citep{dekhtyar2015categorial}.
\item Nearly private information retrieval - This is a research concerning the privacy of data that suggest improvement to the previous approaches for keeping the data retrieval safe and secure \citep{chakrabarti2007nearly}.  
\end{itemize}

The above list and the brief explanations show the theoretical theme that is flowing in these type of research. This, as can be seen below, is different from what ECS targets.

\subsubsection{Experimental Computer Science}

Experimental Computer Science (ECS) is the body of best practices, methods, procedures, and techniques that assist the practitioners of computing in moving the computer science from its theoretical base towards an applied science. Although the computing seems to be an experimental area, research showed that up to 1995 this was not the case, at least by assessing the published results \citep{tichy1995experimental}. However, it does not mean that the theoretical research was a dominant area. In fact, according to \cite{tichy1995experimental}, about 70\% of published papers by ACM (Association for Computing Machinery) was rather design and modeling. A quick search using different search engines are still showing that this situation has continued to some extent, which is confirmed by \cite{wainer2009empirical} in as well. Hence a call for a cultural change in computer science towards showing more appreciation for experimental approaches was still there in 2006 \citep{feitelson2006experimental}.

\par Despite this situation, we can find experimental research of high quality nowadays. Below some samples are mentioned. However, it is still to soon to say that the research in computer science is a well-established discipline. \footnote{The author has experienced the confusion and misunderstanding about research in computing in both industry and academy. It still seems difficult to convince the students to completely differentiate between a software development, for example, with an experimental research in computing. The research methods, in general, seem to be much appreciated and understood by students of other fields of science (social or natural) and engineering rather than computing.}

\begin{itemize}
\item Verification and change-impact analysis of access-control policies - This research investigates the data access-control policies through using a software, which is called Margrave. The aim is to measure how changes in the policies would affect the performance \citep{fisler2005verification}.
\item A two-tier test approach to developing location-aware mobile learning systems
for natural science courses - This research conducts experiments to assess the effectiveness of mobile learning system on elementary school students \citep{chu2010two}.
\end{itemize}

Having considered the two main areas of computer science research, it is also seen that sometime researchers talk about ``emperical'' computer science. By the same analogy that emperical research has been distinguished from experimental research in other sciences, these two are also distinguished in computing and computer science. Nevertheless, emperical research has been given some especial attention in Software Engineering area of computing \citep{perry2000empirical,wohlin2003empirical,easterbrook2008selecting}.    

\section{Summary and Conclusion}

Research is one of the pillars of advancement in science and technology. It is a methodical approach for finding answers to the problems through investigation and experimentations by which researchers evaluate a hypothesis, provide answers to the research questions, or suggest solutions to certain problems. Although the main goal of research is the same for all branches of science and humanities, the characteristics of each branch requires a specific    adaptation of the methods which are applicable for the research. To choose a proper method and to design the way that the research should be carried out, researchers should discuss and assess these methods in the context of the research. This process and its outcome, together, is called research methodology.

\par Computing, in general, and computer science, in particular are relatively new sciences. Although the convergence of different branches of science is a phenomenon in the new millennium, the interdisciplinary feature of computing is far more evident and effective than any other sciences. The ubiquity of computers not only have affected the lifestyle of human being, but also has changed the interconnection between all sectors of science and humanities. On the one hand, this interconnection has led the researchers in computing to borrow and adapt the research methods and methodologies which have been used for a long time in well-established sciences. On the other hand, this has caused the research methodology in computing to remain premature. Furthermore, computing includes two aspects, science and engineering. For this, research methodology of computing is usually discussed in two dimensions, theoretical and experimental.

\par From the paradigmatic perspective, computing research falls in positivism/post-positivism paradigm. Accordingly, although it mainly uses quantitative methods, using qualitative and mixed-methods are also common among researchers; see \citep{wohlin2003empirical}, for example. In fact, sometimes mixed-methods are the best choice for computing research, especially where the research is overlapped with some other branches, for instance, social science areas. 

\par As a result, despite the current confusion on computing research methods and methodology, it seems that its main challenges such as identity, proper adaptation, and established educational methods have received a substantial attention among the computing research community. Consequently, this would hopefully help a well-established computing research method and methodology to appear in the near future. Until then, computing researchers should continue to polish their ideas on research method and methodology, particularly, helping students and novice researchers to differentiate and distinguish the sometimes blurred area between application/software development/system development {\it production projects} and applied/experimental/theoretical {\it research projects}.

\begingroup
\endgroup

\bibliographystyle{chicago}
\bibliography{ResearchMethodsInCS}

\end{document}